\documentclass[aps,prd,nofootinbib,amsmath,amssymb,superscriptaddress,twocolumn,10pt]{revtex4}%superscriptaddress,showpacs,
\usepackage{graphicx}
\usepackage{dcolumn}
\usepackage{bm}
\usepackage{amssymb}
\usepackage{latexsym}
\usepackage{booktabs}
\usepackage{amsmath}
\usepackage{multirow}
\usepackage[colorlinks=true, linkcolor=red, citecolor=blue]{hyperref}

\newcommand{\be}{\begin{equation}}
\newcommand{\ee}{\end{equation}}
\newcommand{\bq}{\begin{eqnarray}}
\newcommand{\eq}{\end{eqnarray}}

\begin{document}

\title{Measuring growth index in a universe with sterile neutrinos}

\author{Jing-Fei Zhang}
\affiliation{Department of Physics, College of Sciences, Northeastern University, Shenyang
110004, China}
\author{Yun-He Li}
\affiliation{Department of Physics, College of Sciences, Northeastern University, Shenyang
110004, China}
\author{Xin Zhang\footnote{Corresponding author}}
\email{zhangxin@mail.neu.edu.cn} \affiliation{Department of Physics, College of Sciences,
Northeastern University, Shenyang 110004, China}
\affiliation{Center for High Energy Physics, Peking University, Beijing 100080, China}

\begin{abstract}
Consistency tests for the general relativity (GR) can be performed by constraining the growth index $\gamma$ using the 
measurements of redshift-space distortions (RSD) in conjunction with other observations. In previous studies, deviations from 
the GR expected value of $\gamma\approx 0.55$ at the 2--3$\sigma$ level were found. In this work, we reconsider the measurement of 
$\gamma$ in a universe with sterile neutrinos. We constrain the sterile neutrino cosmological model using the RSD measurements combined 
with the cosmic microwave background data (Planck temperature data plus WMAP 9-yr polarization data), the baryon acoustic oscillation data, 
the Hubble constant direct measurement, the Planck Sunyaev-Zeldovich cluster counts data, and the galaxy shear data. 
We obtain the constraint result of the growth index, $\gamma=0.584^{+0.047}_{-0.048}$, well consistent with the GR expected value
(the consistency is at the 0.6$\sigma$ level). For the parameters of sterile neutrino, we obtain $N_{\rm{eff}}=3.62^{+0.26}_{-0.42}$ and 
$m_{\nu,{\rm{sterile}}}^{\rm{eff}}=0.48^{+0.11}_{-0.14}$ eV. We also consider the BICEP2 data and perform an analysis on the model 
with tensor modes. Similar fit results are obtained, showing that once light sterile neutrino is considered in the universe, 
GR will become well consistent with the current observations. 

\end{abstract}

\pacs{95.36.+x, 98.80.Es, 98.80.-k} \maketitle

%\section{Introduction}

Since the discovery of the acceleration of the current universe's expansion, 
dark energy has been viewed as the mainstream for explaining the cause of this phenomenon \cite{de1,de2}. 
In particular, the cosmological constant ($\Lambda$) plus cold dark matter (CDM) model, called the 
$\Lambda$CDM model, has been achieving successes in fitting various observational data. 
Notwithstanding, other possibilities explaining the cosmic acceleration still exist, among which the most 
popular alternative is the modification to general relativity (GR).
Various modified gravity (MG) models in the large-scale/weak-field limit have been proposed; 
for recent reviews, see, e.g., Refs. \cite{mg1,mg2}. 
Since dark energy and MG theories can in principle predict identical expansion histories, a potential way of 
distinguishing between them is to probe and compare the different structure growth histories of them. 

In linear perturbation theory, it is possible to describe the cosmic structure's growth history through a second-order 
differential equation, which depends on both the Hubble expansion rate $H(z)$ and the specific theory of gravity. 
Solving this differential equation, one can derive the dimensionless linear growth rate, $f(a)=d\ln D(a)/d\ln a$, 
describing how rapidly structure grows as a function of cosmic scale factor $a(t)$, where $D(a)$ is the growth factor 
depicting the growth of the matter perturbations at late times. 
A fitting formula of $f(a)=\Omega_m(a)^\gamma$ proposed by Wang and Steinhardt \cite{Wang:1998gt} has been proven to be an accurate description 
for a wide range of models \cite{Linder:2005in}, in which both the growth index $\gamma$ and the fractional matter density $\Omega_m(a)=\Omega_m 
H_0^2 H(a)^{-2}a^{-3}$ depend on the specific model. 
For dark energy models with slowly varying equation of state (within the framework of GR), $\gamma$ was analytically given, 
showing that $\gamma=6/11\approx 0.545$ for the $\Lambda$CDM model and actually $\gamma\approx 0.55$ corresponds to 
a wide range of dark energy models in GR \cite{Wang:1998gt,Linder:2005in}. For the MG models, different values of $\gamma$ can be derived; for example, 
$\gamma\approx 0.68$ is obtained theoretically for the Dvali-Gabadadze-Porrati (DGP) braneworld model \cite{Lue:2004rj}. 

Redshift-space distortions (RSD) provide an important way of measuring $f(z)$ at different redshifts \cite{Peacock:2001gs,Guzzo:2008ac}. 
RSD arise from peculiar velocities of galaxies on observed galaxy map. 
Since the coherent motions of galaxies are actually a direct consequence of the growth of structure, 
the measurement of anisotropy they induce in the redshift space provides information about the formation of large-scale structure. 
In practice, RSD measure the combination of $f(z)$ and $\sigma_8(z)$, i.e., $f(a)\sigma_8(a)=d\sigma_8(a)/d\ln a$, 
where $\sigma_8(z)$ is the root-mean-square mass fluctuation in spheres with radius $8h^{-1}$ Mpc at redshift $z$ \cite{Song:2008qt}. 

Consistency tests of GR using the RSD measurements have been performed in the literature. 
In Ref.~\cite{Samushia:2012iq}, Samushia et al. used the BOSS CMASS DR9 measurement of growth rate in combination with 
cosmic microwave background (CMB) and type Ia supernova (SN) data to constrain the growth index of 
the $\Lambda$CDM model and obtained $\gamma=0.75\pm 0.09$; 
when other $f\sigma_8$ measurements are added (totally 9 data), the result is improved to 
$\gamma=0.64\pm 0.05$. In both cases, the constraint results are in tension with the GR expected value of 
$\gamma=0.55$ at about the 2$\sigma$ level. 
In Ref.~\cite{Beutler:2013yhm}, Beutler et al. used the BOSS CMASS DR11 data (including $D_V/r_s$, $F_{AP}$, and $f\sigma_8$ at the effective redshift $z=0.57$) 
combined with the Planck data to place constraint on the growth index and obtained the result $\gamma=0.772^{+0.124}_{-0.097}$, 
in tension with the GR expected result at the 2.3$\sigma$ level.  
Similar situation can also be found in other studies, e.g., Ref.~\cite{Xu:2013tsa}, in which the constraint result of $\gamma$ from the RSD measurements (10 $f\sigma_8$ data in total) 
combined with the SN, baryon acoustic oscillation (BAO), and Planck data is also discrepant from the GR expected value at about the 2--3$\sigma$ level.

In this paper, we consider the growth rate of structure in a universe with sterile neutrinos. 
Considering the parametrization $f(z)=\Omega_m(z)^\gamma$, we make a consistency test for GR in this case. 
In our previous work \cite{zx14s}, we have shown that involving a light (eV or sub-eV mass scale) sterile neutrino species in the $\Lambda$CDM model can help 
reconcile the tensions between Planck and other astrophysical observations (such as the direct measurement of $H_0$ of HST observation, the 
Planck Sunyaev-Zeldovich cluster counts, the cosmic shear measurement of CFHTLenS survey, and the CMB polarization measurements of BICEP2). 
It was shown that, under the CMB+BAO constraint, the inconsistencies with Planck are improved from 2.4$\sigma$ to 1.0$\sigma$ for $H_0$ observation, 
from 4.3$\sigma$ to 2.0$\sigma$ for SZ cluster counts observation, and from 2.3$\sigma$ to 1.7$\sigma$ for cosmic shear observation; at the same time, 
the 95\% upper limit for the tensor-to-scalar ratio $r_{0.002}$ is also enhanced to 0.20, relieving the tension between Planck and BICEP2 \cite{zx14s}. 
Furthermore, combining CMB data with growth of structure measurements could provide independent evidence for the existence of light sterile neutrinos, i.e., 
the current cosmological data prefer $\Delta N_{\rm eff}>0$ at the 2.7$\sigma$ level and a nonzero mass of sterile neutrino at the 3.9$\sigma$ level \cite{zx14s}. 
For other relevant studies, see, e.g., 
Refs.~\cite{hu14,zx14s2,zx14s3,zx14s4, nus1,nus2,sterile1,sterile2,sterile3}. 
Here we caution the reader that the systematic errors in the astrophysical observations, such as cluster counts, weak lensing, and Hubble constant measurements, could weaken these conclusions to some extent. 
In particular, it was shown in Ref.~\cite{Leistedt:2014sia} that the standard $\Lambda$CDM model is favored over the model with sterile neutrino if the Bayesian evidence is used as a criterion for comparing models.

Since the growth of structure could be suppressed by light sterile neutrinos in the $\Lambda$CDM cosmology, 
perhaps the growth index constrained by the RSD measurements would become consistent with the GR expected value of $\gamma\approx 0.55$. 
This motivates the present work. In the following, we test GR using the RSD measurements (in conjunction with other observations) in the universe with sterile neutrinos.

%\section{Observational data}

%\subsection{Redshift space distortions}

We use the recent RSD measurements to constrain the growth index $\gamma$. The $f(z)\sigma_8(z)$ data we consider in this work include 
the measurements from 6dFGS ($z=0.067$) \cite{RSD6dF}, 2dFGRS ($z=0.17$) \cite{RSD2dF}, WiggleZ ($z=0.22$, 0.41, 0.60, and 0.78) \cite{RSDwigglez}, 
SDSS LRG DR7 ($z=0.25$ and 0.37) \cite{RSDsdss7}, BOSS CMASS DR11 ($z=0.57$) \cite{Beutler:2013yhm}, and VIPERS ($z=0.80$) \cite{RSDvipers}. 
In Ref.~\cite{Beutler:2014yhv}, through a test of the reliability of low redshift growth of structure measurements (including RSD), 
it was shown that the cosmological parameter constraints are robust against changes in the power spectrum template related to neutrino mass. 
This justifies the use of the RSD data as well as other growth of structure data in this work.

%\subsection{Other data sets}

To constrain other parameters and break degeneracies between parameters, we should also employ other observational data. 
We consider the CMB, BAO, $H_0$, cluster counts, and cosmic shear data. 
For the CMB data, we use the Planck temperature power spectrum data \cite{planck} in combination with the WMAP 9-yr polarization (TE and EE) power spectrum data \cite{wmap9}. 
For the BAO data, we use the measurements from 6dFGS ($z=0.1$) \cite{6df}, SDSS DR7 ($z=0.35$) \cite{sdss7}, 
WiggleZ ($z=0.44$, 0.60, and 0.73) \cite{wigglez}, and BOSS DR11 ($z=0.32$ and 0.57) \cite{boss}; 
note that the three data from WiggleZ survey are correlated, with the inverse covariance matrix given in Ref.~\cite{wigglez}. 
For the $H_0$ measurement, we use the HST result, $H_0=(73.8\pm 2.4)$ km s$^{-1}$ Mpc$^{-1}$ \cite{h0}. 
For the Planck SZ (PlaSZ) measurement, we use the result of $\sigma_8(\Omega_m/0.27)^{0.3}=0.764\pm 0.025$ (which is derived by allowing the bias $(1-b)$ 
to vary in the range of $[0.7, 1]$) \cite{tsz}. For the shear measurement, we use the CFHTLenS result, $\sigma_8(\Omega_m/0.27)^{0.6}=0.79\pm 0.03$ \cite{shear}. 
The consistency of these data sets has been tested for the universe with sterile neutrinos, and so the combination of these data 
is appropriate \cite{zx14s,zx14s4}.

%\section{Results and discussions}

%\subsection{Growth index and sterile neutrinos}

%===================Table 1===========================================
\begin{table}[tbp]
\centering\caption{\label{table1} Constraint results for the $\Lambda$CDM, $\Lambda$CDM+$\nu_s$, and
$\Lambda$CDM+$r$+$\nu_s$ models. Note that the mass of sterile neutrino $m_{\nu,{\rm{sterile}}}^{\rm{eff}}$
is in unit of eV and the Hubble constant $H_0$ is in unit of km s$^{-1}$ Mpc$^{-1}$.}
\begin{tabular}{lccc}
\hline
\hline
Parameter &$\Lambda$CDM& $\Lambda$CDM+$\nu_s$  & $\Lambda$CDM+$r$+$\nu_s$ \\
\hline
$\gamma$&$0.667^{+0.049}_{-0.053}$&$0.584^{+0.047}_{-0.048}$&$0.600^{+0.044}_{-0.049}$\\
$N_{\rm{eff}}$&...&$3.62^{+0.26}_{-0.42}$&$4.01^{+0.30}_{-0.33}$\\
$m_{\nu,{\rm{sterile}}}^{\rm{eff}}$&...&$0.48^{+0.11}_{-0.14}$&$0.53^{+0.11}_{-0.14}$\\
$r_{0.002}$&...&...&$0.221^{+0.042}_{-0.052}$\\
$\Omega_m$&$0 .3051^{+0.0073}_{-0.0072}$&$0.3015^{+0.0076}_{-0.0075}$&$0.3009^{+0.0073}_{-0.0080}$\\
$\sigma_8$&$0.822\pm0.011$&$0.746\pm0.013$&$0.745^{+0.013}_{-0.012}$\\
$H_0$&$68.0\pm0.6$&$70.5^{+1.3}_{-1.7}$&$72.0\pm1.3$\\
\hline
$-\ln\mathcal{L}_{\rm{max}}$ &4908.591&4912.801 &4932.818   \\
\hline\hline
\end{tabular}
\end{table}
%====================================================================

%\begin{table}[tbp]
%\centering\caption{\label{table1} old SZ+shear and WL}
%\begin{tabular}{lcc}
%\hline
%\hline
%Parameters & old  & WL \\
%\hline
%$\gamma$&$0.598\pm0.046$&$0.582^{+0.049}_{-0.051}$\\
%$N_{\rm{eff}}$&$3.65^{+0.28}_{-0.37}$&$3.60^{+0.24}_{-0.46}$\\
%$m_{\nu,{\rm{sterile}}}^{\rm{eff}}$&$0.42^{+0.10}_{-0.12}$&$0.49^{+0.12}_{-0.19}$\\
%\hline\hline
%\end{tabular}
%\end{table}

%=======================Figure 1======================================
\begin{figure}[tbp]
\centering % \begin{center}/\end{center} takes some additional vertical space
\includegraphics[width=7.3cm]{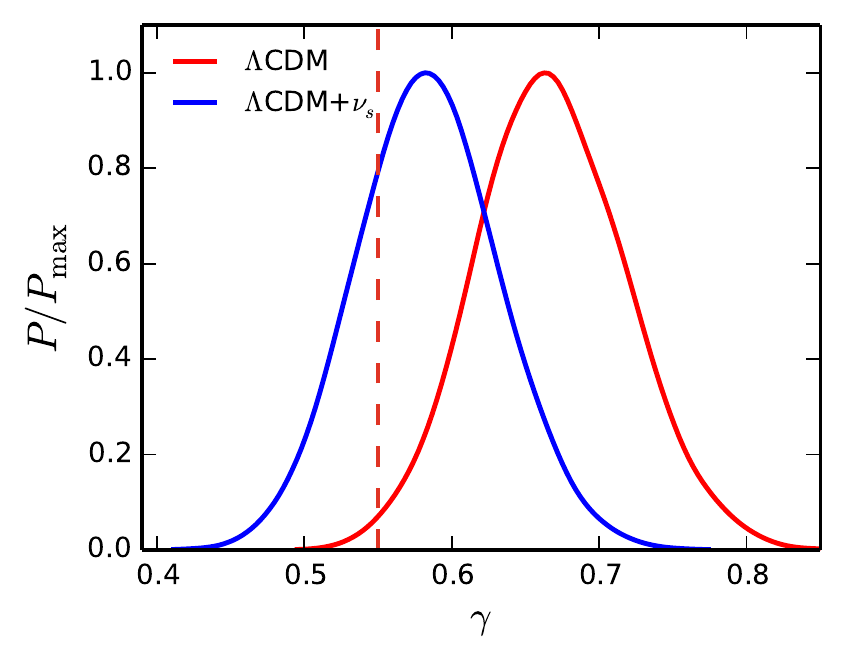}
\includegraphics[width=7.5cm]{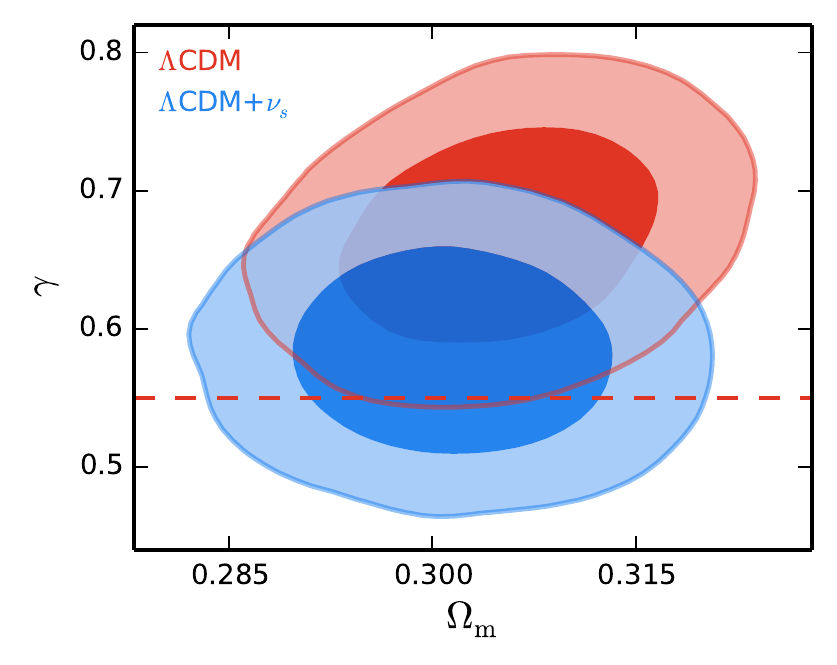}
\hfill
%\includegraphics[width=.45\textwidth,origin=c,angle=180]{img2.pdf}
% "\includegraphics" is very powerful; the graphicx package is already loaded
\caption{\label{fig1} Constraint results for the $\Lambda$CDM+$\nu_s$ model from the CMB+BAO+$H_0$+PlaSZ+Shear+RSD data combination. 
Upper panel: one-dimensional marginalized posterior distribution for the growth index $\gamma$. 
Lower panel: two-dimensional marginalized posterior distribution contours (68.3\% and 95.4\% CL) in the $\Omega_m$--$\gamma$ plane. 
The constraint results of the $\Lambda$CDM model (without sterile neutrino) using the CMB+BAO+RSD data 
are also shown for a comparison with the model with sterile neutrino.
The vertical (upper) and horizontal (lower) red dashed lines indicate the GR expected value of $\gamma=0.55$.}
\end{figure}
%=====================================================================

%========================Figure 2======================================
\begin{figure}[tbp]
\centering % \begin{center}/\end{center} takes some additional vertical space
\includegraphics[width=7.5cm]{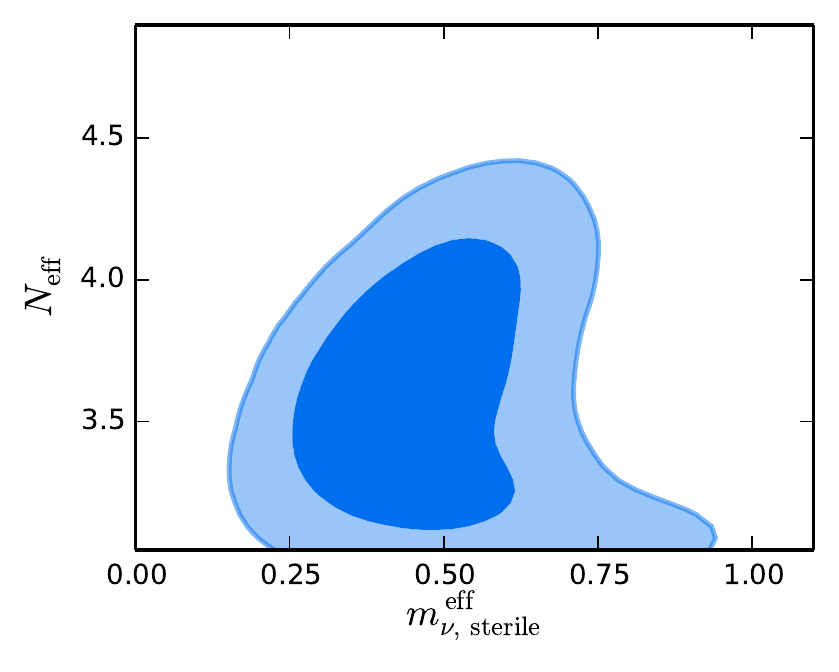}
\hfill
%\includegraphics[width=.45\textwidth,origin=c,angle=180]{img2.pdf}
% "\includegraphics" is very powerful; the graphicx package is already loaded
\caption{\label{fig2} Two-dimensional marginalized constraints (68.3\% and 95.4\% CL) on the $\Lambda$CDM+$\nu_s$ model 
from the CMB+BAO+$H_0$+PlaSZ+Shear+RSD data combination in the $m_{\nu,{\rm{sterile}}}^{\rm{eff}}$--$N_{\rm{eff}}$ plane.}
\end{figure}
%====================================================================

Following our previous work \cite{zx14s}, the model with sterile neutrinos considered in the paper is called 
the $\Lambda$CDM+$\nu_s$ model, in which the active neutrino mass is kept at 0.06 eV (minimal-mass normal hierarchy is assumed) and two additional parameters 
related to sterile neutrino, $N_{\rm{eff}}$ and $m_{\nu,{\rm{sterile}}}^{\rm{eff}}$, are involved. In addition, the growth index $\gamma$ is introduced through the parametrization 
of the growth rate $f(z)=\Omega_m(z)^\gamma$. To place constraints on $\gamma$, we follow the recipe of Ref.~\cite{Beutler:2013yhm} to calculate the values of $f(z)\sigma_8(z)$ in the 
numerical code (we modify the {\tt CosmoMC} code \cite{cosmomc}). In the following, we report the results of the parameter estimation.

We use the CMB+BAO+$H_0$+PlaSZ+Shear+RSD data combination to constrain the $\Lambda$CDM+$\nu_s$ model. The main constraint 
results are shown in Table~\ref{table1} as well as Figs.~\ref{fig1} and \ref{fig2}. 
We obtain $\gamma=0.584^{+0.047}_{-0.048}$, well consistent with the GR expected value of $\gamma=0.55$. 
Figure~\ref{fig1} shows the one-dimensional posterior distribution for $\gamma$ (upper panel) and two-dimensional posterior contours 
for $\gamma$ and $\Omega_m$ (lower panel). 
One can clearly see that GR lies well inside the 1$\sigma$ range; the consistency is at the 0.6$\sigma$ level.  
In addition, the parameters of sterile neutrino can also be precisely determined in this case. 
The constraint result for sterile neutrino in the $m_{\nu,{\rm{sterile}}}^{\rm{eff}}$--$N_{\rm{eff}}$ plane is shown in Fig.~\ref{fig2}.
We obtain $N_{\rm{eff}}=3.62^{+0.26}_{-0.42}$ and $m_{\nu,{\rm{sterile}}}^{\rm{eff}}=0.48^{+0.11}_{-0.14}$ eV, 
indicating the preference for $\Delta N_{\rm eff}\equiv N_{\rm eff}-3.046>0$ at the 1.4$\sigma$ level and for nonzero mass of 
sterile neutrino at the 3.4$\sigma$ level. 

We also show the constraint results of the $\Lambda$CDM model (without 
sterile neutrino) using the CMB+BAO+RSD data combination in Fig.~\ref{fig1}, 
for a direct comparison with the model with sterile neutrino. 
Note that here we do not incorporate the $H_0$, PlaSZ and Shear data in the analysis, 
since there exist known tensions between Planck data and these three measurements 
for the $\Lambda$CDM model. 
For this case, we obtain $\gamma=0.667^{+0.049}_{-0.053}$, discrepant from the GR expected value 
at the 2.3$\sigma$ level. The fitting results for this case are also given in Table~\ref{table1}. 
From Fig.~\ref{fig1}, we can directly see what a crucial role the sterile neutrino plays in 
changing the fit value of $\gamma$, leading to the consistency with GR.

%The inclusion of massive sterile neutrinos is important for reducing tension between Planck and large-scale structure observations, 
%as indicated in Refs.~\cite{zx14s,hu14,zx14s2,zx14s3,zx14s4, sterile1,sterile2,sterile3}, 

The tension with GR is mainly caused by the large $\sigma_8$ in the $\Lambda$CDM model fitting to the 
CMB data. 
When light massive sterile neutrino is involved in the model, the growth of structure could be suppressed below its free streaming length, 
allowing $\sigma_8$ to be substantially lower. 
This explains why the tension with GR is reduced (i.e., the RSD constraint on the growth index $\gamma$ becomes in good agreement with the GR prediction, as shown in the above analysis), once a light sterile neutrino is introduced into the model. Moreover, the same reason leads to the reduction of tensions between Planck and other large-scale structure observations, as indicated in Refs.~\cite{zx14s,hu14,zx14s2,zx14s3,zx14s4,nus1,nus2, sterile1,sterile2,sterile3}.

%This leads to the reduction of tension between Planck and large-scale structure observations, 
%as indicated in Refs.~\cite{zx14s,hu14,zx14s2,zx14s3,zx14s4,nus1,nus2, sterile1,sterile2,sterile3}. 
%Also, light sterile neutrino could make the constraint on the growth index $\gamma$ in good agreement with the GR prediction, as shown 
%in the above analysis. 
%The same reason explains why the tension with GR is reduced once a light sterile neutrino is introduced into the model. 

Since the cluster counts and weak lensing data are important for determining the properties of sterile neutrino, we 
make a further analysis. In previous studies, e.g., \cite{zx14s,zx14s2,zx14s3,zx14s4, nus1,sterile1,sterile3}, somewhat different PlaSZ and Shear data 
were used, i.e., $\sigma_8(\Omega_m/0.27)^{0.3}=0.782\pm 0.010$ (derived by fixing $1-b=0.8$) \cite{tsz} and $\sigma_8(\Omega_m/0.27)^{0.46}=0.774\pm 0.040$ \cite{shear2}.
Replacing with these two measurement results in our analysis, we obtain the constraint results: 
$\gamma=0.598\pm0.046$, $N_{\rm{eff}}=3.65^{+0.28}_{-0.37}$, and $m_{\nu,{\rm{sterile}}}^{\rm{eff}}=0.42^{+0.10}_{-0.12}$ eV. 
In addition, due to the existence of some unknown systematic uncertainties in the PlaSZ data, we also perform an analysis without PlaSZ. 
We follow Ref.~\cite{Beutler:2014yhv} to replace PlaSZ with the galaxy-galaxy lensing result, $\sigma_8(\Omega_m/0.25)^{0.57}=0.80\pm 0.05$ \cite{Mandelbaum:2012ay}, 
and redo the analysis. In this case, we obtain the fit results: $\gamma=0.582^{+0.049}_{-0.051}$, $N_{\rm{eff}}=3.60^{+0.24}_{-0.46}$, and 
$m_{\nu,{\rm{sterile}}}^{\rm{eff}}=0.49^{+0.12}_{-0.19}$ eV. Comparing the three cases, we find that their results are in good agreement with each other.

%\subsection{With tensor modes and BICEP2 data}

%=======================Figure 3======================================
\begin{figure}[tbp]
\centering % \begin{center}/\end{center} takes some additional vertical space
\includegraphics[width=7.5cm]{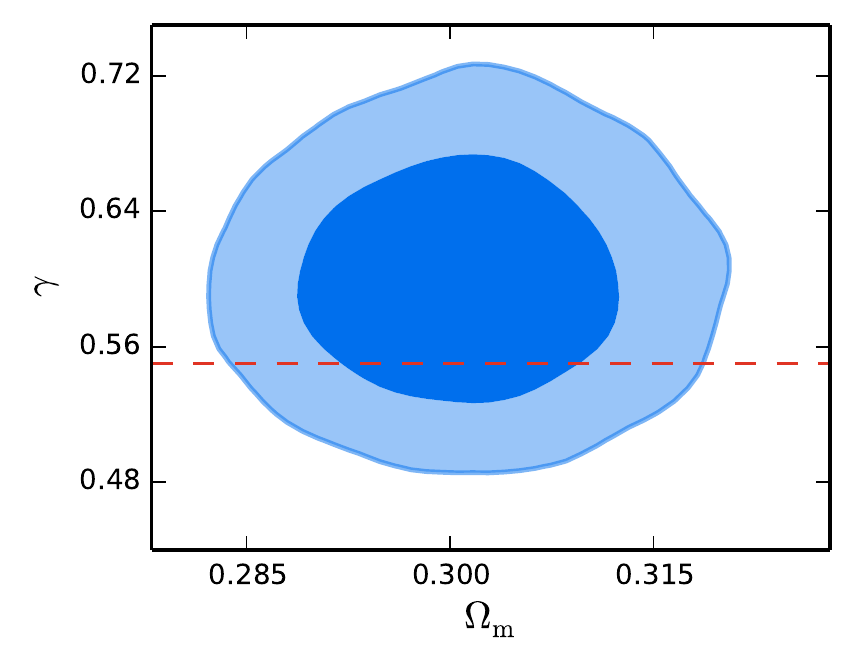}
\includegraphics[width=7.5cm]{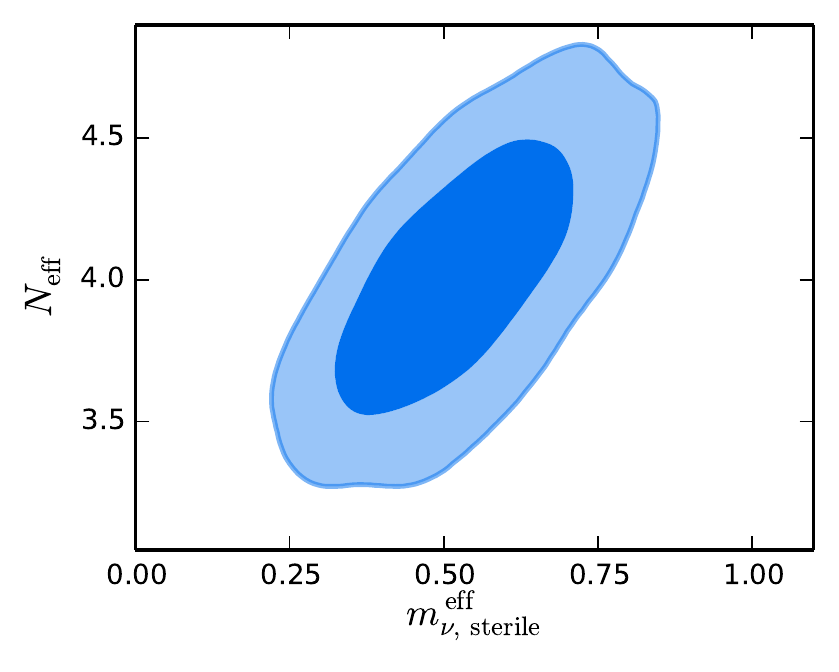}
\hfill
%\includegraphics[width=.45\textwidth,origin=c,angle=180]{img2.pdf}
% "\includegraphics" is very powerful; the graphicx package is already loaded
\caption{\label{fig3} Joint constraint results (68.3\% and 95.4\% CL) for the $\Lambda$CDM+$r$+$\nu_s$ model from the CMB+BAO+$H_0$+PlaSZ+Shear+RSD+BICEP2 data combination 
in the $\Omega_m$--$\gamma$ plane (upper) and in the $m_{\nu,{\rm{sterile}}}^{\rm{eff}}$--$N_{\rm{eff}}$ plane (lower). 
The horizontal red dashed line in the upper panel indicates the GR expected value of $\gamma=0.55$.}
\end{figure}
%===================================================================

Recently, the detection of the B-mode polarization of CMB was reported by the BICEP2 Collaboration \cite{bicep2}. 
Provided that the foreground model was treated correctly, the BICEP2's result would indicate the discovery of the primordial gravitational waves (PGWs). 
If confirmed by upcoming experiments, the frontiers of fundamental physics will be pushed forward in an unprecedented way. 
Fitting to the BICEP2 data gives an unexpectedly large tensor-to-scalar ratio, $r=0.20^{+0.07}_{-0.05}$, in tension with the 95\% upper limit, $r<0.11$, 
given by Planck. In our previous work \cite{zx14s}, we have shown that involving sterile neutrino in the model (i.e., considering the $\Lambda$CDM+$r$+$\nu_s$ model) 
could well relieve the tension between Planck and BICEP2, and meanwhile, the other tensions of Planck with other astrophysical observations can all be significantly reduced
(see also Ref.~\cite{hu14}). 

In the present work, we also perform an analysis for the $\Lambda$CDM+$r$+$\nu_s$ model. 
In this case, we also use the BICEP2 data  \cite{bicep2} to carry out the joint constraints. 
The main constraint results are shown in Table~\ref{table1} and Fig.~\ref{fig3}. 
The fit result for the growth index is $\gamma=0.600^{+0.044}_{-0.049}$, also consistent with the GR prediction of $\gamma=0.55$ at the 1$\sigma$ level 
(see the upper panel of Fig.~\ref{fig3}). 
For the parameters of sterile neutrino, we obtain $N_{\rm{eff}}=4.01^{+0.30}_{-0.33}$ and $m_{\nu,{\rm{sterile}}}^{\rm{eff}}=0.53^{+0.11}_{-0.14}$ eV 
(see the lower panel of Fig.~\ref{fig3}), indicating the preference for $\Delta N_{\rm eff}>0$ at the 2.9$\sigma$ level and for nonzero mass of sterile neutrino 
at the 3.8$\sigma$ level.

%\section{Conclusion}

In summary, we have performed a consistency test for GR in a universe with sterile neutrinos through constraining the growth index $\gamma$ 
using the RSD data in conjunction with other observations. 
The observational data we used in this work include the CMB, BAO, $H_0$, PlaSZ, Shear, and RSD data. 
For the $\Lambda$CDM+$\nu_s$ model, we obtained the result of $\gamma=0.584^{+0.047}_{-0.048}$, well consistent with the GR expected value of $\gamma=0.55$; 
the consistency is at the 0.6$\sigma$ level. 
We obtained the parameters of sterile neutrino, $N_{\rm{eff}}=3.62^{+0.26}_{-0.42}$ and $m_{\nu,{\rm{sterile}}}^{\rm{eff}}=0.48^{+0.11}_{-0.14}$ eV, 
indicating the preference for $\Delta N_{\rm eff}>0$ at the 1.4$\sigma$ level and for nonzero mass of 
sterile neutrino at the 3.4$\sigma$ level.
We also tested the $\Lambda$CDM+$r$+$\nu_s$ model, and the BICEP2 data were also included in the analysis for this case. 
The constraint results are $\gamma=0.600^{+0.044}_{-0.049}$,  $N_{\rm{eff}}=4.01^{+0.30}_{-0.33}$, and $m_{\nu,{\rm{sterile}}}^{\rm{eff}}=0.53^{+0.11}_{-0.14}$ eV, 
indicating the consistency with GR at the 1$\sigma$ level as well as the preference for $\Delta N_{\rm eff}>0$ at the 2.9$\sigma$ level and for nonzero mass of sterile neutrino 
at the 3.8$\sigma$ level. The previous consistency tests for GR in the literature show some tension at about the 2--3$\sigma$ level. 
In this work, we show that when light sterile neutrinos are considered in the universe, GR will become well consistent with the current observations.

\begin{acknowledgments}
We acknowledge the use of {\tt CosmoMC}. 
JFZ is supported by the Provincial Department of Education of
Liaoning under Grant No. L2012087.
XZ is supported by the National Natural Science Foundation of
China under Grant No. 11175042 and the Fundamental Funds for the 
Central Universities under Grant No. N120505003.
\end{acknowledgments}

\end{document}